\documentclass[12pt]{article}
\newwrite\ffile\global\newcount\figno \global\figno=1

\def\writedef#1{}

\input epsf
\def\figin{\epsfcheck\figin}\def\figins{\epsfcheck\figins}
\def\epsfcheck{\ifx\epsfbox\UnDeFiNeD
\message{(NO epsf.tex, FIGURES WILL BE IGNORED)}
\gdef\figin##1{\vskip2in}\gdef\figins##1{\hskip.5in}
\else\message{(FIGURES WILL BE INCLUDED)}%
\gdef\figin##1{##1}\gdef\figins##1{##1}\fi}

\def\figinsert{}
\def\ifig#1#2#3{\xdef#1{fig.~\the\figno}
\writedef{#1\leftbracket fig.\noexpand~\the\figno}%
\figinsert\figin{\centerline{#3}}\medskip\centerline{\vbox{\baselineskip12pt
\advance\hsize by -1truein\center\footnotesize{  Fig.~\the\figno.} #2}}
\bigskip\endinsert\global\advance\figno by1}
\def\endinsert{}

\usepackage{amssymb}
\begin{document}
\baselineskip 18pt
\newcommand{\Tr}{\mbox{Tr\,}}
\newcommand{\beq}{\begin{equation}}
\newcommand{\eeq}{\end{equation}}
\newcommand{\bea}{\begin{eqnarray}}
\newcommand{\eea}[1]{\label{#1}\end{eqnarray}}
\renewcommand{\Re}{\mbox{Re}\,}
\renewcommand{\Im}{\mbox{Im}\,}

\def\N{{\cal N}}


\thispagestyle{empty}
\renewcommand{\thefootnote}{\fnsymbol{footnote}}

{\hfill \parbox{4cm}{
        SHEP-04-19 \\
}}

\bigskip

\begin{center} \noindent {\Large \bf
Higgsless W Unitarity from Decoupling Deconstruction }
\end{center}

\bigskip\bigskip\bigskip

\centerline{ \normalsize \bf Nick Evans and Phil Membry
\footnote[1]{\noindent \tt evans@phys.soton.ac.uk,
pjm@phys.soton.ac.uk}}

\bigskip
\bigskip\bigskip

\centerline{\it School of Physics and Astronomy}
\centerline{ \it
Southampton University} \centerline{\it Southampton, S017
1BJ }
\centerline{ \it United Kingdom}
\bigskip

\bigskip\bigskip

\renewcommand{\thefootnote}{\arabic{footnote}}

\centerline{\bf \small Abstract} Recently there has been
interest
in electroweak models on a five dimensional interval that
break
the symmetry without a higgs boson. By warping the metric
of the
interval it may be possible to avoid experimental bounds on
extra
W bosons and $\delta \rho$. Five dimensional models
necessarily
require an explicit UV cut-off to remain perturbative, such
as
that provided by deconstruction. We study a simple
deconstruction
of this scenario with a chain of $SU(2)^{N+1} \times
U(1)_Y$
groups linked by bi-fundamental higgs. There are two
interesting
decoupling limits of this model, when the higgs vevs are
taken
large and when the SU(2) couplings grow, which might
provide a
perturbative realization. In fact it is very challenging to
satisfy all the experimental bounds and the most compatible
scenario has both a higgs and a relatively strongly coupled
new W
both close to 2 TeV.
\medskip

{\small \noindent }

\newpage


The need to restore unitarity in high energy $WW$
scattering has
long been cited as evidence that there must be a higgs
boson with
mass below of order 1 TeV \cite{olduni}. Recently though it has been
realized
that unitarity can also be restored by a Kaluza Klein (KK) like
tower
of massive W-bosons without a higgs 
\cite{sekhar1,sekhar2,csaki1,csaki2}. These models 
\cite{csaki1}-\cite{nagasawa}are
variants on
the idea that there is a fifth dimension that is a discrete
interval. The gauge group is broken by boundary conditions
at the
ends of the interval rather than by a higgs mechanism. In
the four
dimensional theory at long distance scales there are only
the W, Z
fields and their KK towers, yet the theory is
unitary.

Such a model must though meet the stringent experimental
constraints on the masses of extra W bosons, and on the
precision
data for $\sin \theta_W$ and $\delta \rho$ (or equivalently
the
parameters $S$ and $T$). Some of the above models in which the extra
dimension is warped have made progress in meeting
these
constraints. However a five dimensional theory is
necessarily ill
defined in the UV where it becomes strongly coupled and
must have
some UV completion before strong coupling is reached. One
must be
careful not to make use of spacetime curvature on scales
where the
theory is strongly coupled - the AdS metric used in \cite{csaki2} with
an
exponential warp factor may for example be hard to support.
Also the
analysis of \cite{schmidt}, which uses the models we study below,
explicitly
works in the strong coupling limit.

To keep track of the gauge coupling strength it is useful
to have
a fully defined theory with an explicit UV completion.
Deconstruction \cite{decon1,decon2} provides such a realization with the fifth
dimension manually constructed by the reproduction of the
Kaluza
Klein tower in a renormalizable four dimensional theory.
 The
extra fifth dimension is first thought of as a lattice 
where a
separate copy of the four dimensional gauge group lives at
each
site. The sites are then linked by Goldstone fields
transforming
in the $(N, \bar{N})$ representation of the two
neighbouring site
gauge groups. The resulting gauge boson mass spectrum, in
the
purely four dimensional model, then mimics a KK
tower at
scales well below the symmetry breaking scale. A fully
renormalizable gauge theory can be found by promoting the
Goldstone fields to a full higgs multiplet. $WW$ unitarity
is
restored by the Kaluza Klein tower at low energies and
finally at
the very high fundamental symmetry breaking scale by the
higgs
bosons \cite{sekhar1}. These models therefore are only higgsless in the
sense
that the higgs mass rises relative to that of the Standard
Model
and phenomenology may appear higgsless at the LHC.

The simplest deconstruction extension of the Standard Model
has
been suggested by a number of authors \cite{schmidt, kurachi, simmons, hirn}. 
It consists of
multiple
repeats of the SU(2) gauge group as shown in the moose
diagram notation \cite{moose1, moose2} of Fig 1. 
There are $N+1$ copies of SU(2) each
potentially with a unique coupling $g_i$. The gauge bosons
are
coupled by bi-fundamental higgs with vevs $v_i$ linking SU(2)$_i$ and
SU(2)$_{i+1}$. Finally the $(N+1)$th SU(2) is coupled by
the
$(N+1)$th higgs to a U(1)$_Y$ hypercharge group. This final
symmetry breaking pattern ensures that there is a
massless photon.

\begin{center}
\hskip-10pt{\lower15pt\hbox{ \epsfysize=1.1 truein
\epsfbox{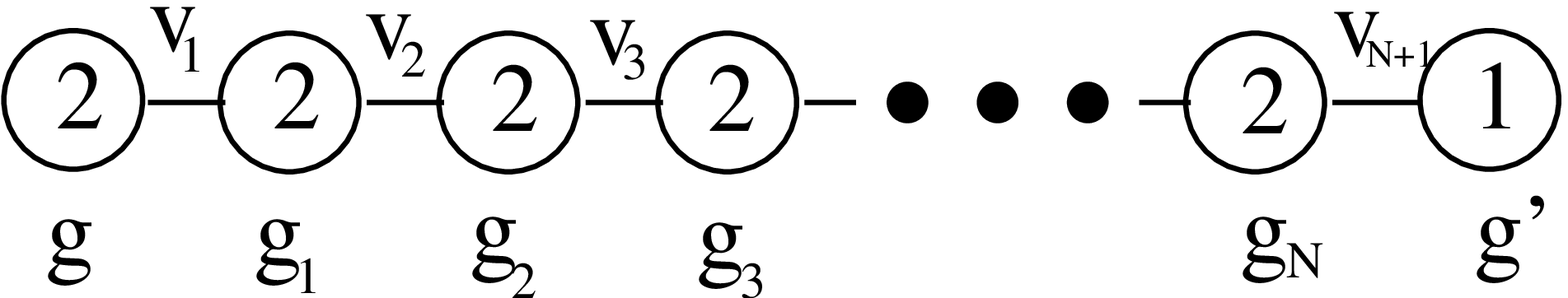}}} \medskip

Figure 1: The moose model under consideration - numbered circles represent
SU(N) gauge groups and links bi-fundamental higgs fields.
\end{center}

The low energy dynamics is described by a non-linear
realization
of the Goldstone fields

\beq {\cal L} = \sum_i {v_i^2 \over 4} Tr D^\mu U_i^\dagger D_\mu
U_i + {\rm higher}~{\rm derivative} \eeq where as usual $U_i =
exp(2 i \pi^a_i T^a/v_i)$ with $\pi^a_i$ the Goldstone fields
associated with the broken generators $T^a$. The gauge fields
enter the covariant derivatives with generators acting on $U_i$
from the left or right depending upon their coupling to the left
or right in the moose diagram.

The tree level W and Z mass matrices may be read off as

{\footnotesize \beq \label{W} M^2_{Wij} = \left( \begin{array}{cccccccc}
g^2 v_1^2 &-g g_1 v_1^2 &0&0&...&0&0&0\\ -g g_1 v_1^2& g_1^2
(v_1^2+v_2^2) & - g_1 g_2 v_2^2 &0 &...&0 &0&0\\ 0 & -g_1 g_2
v_2^2 & - g_2^2(v_2^2 + v_3^2) &- g_2 g_3 v_3^2 &...&0 &0&0\\   .
& . & . &. &...&. &.&.\\  . & . & . &. &...&. &.&.\\   0 & 0 & 0
&0 &...& 0 & -g_{N-1} g_N v_{N-1}^2  & g_N^2 (v_{N-1}^2 + v_N^2)\\
\end{array} \right)\eeq}

{\footnotesize \beq \label{Z} M^2_{Zij} = \left( \begin{array}{ccccccccc}
g^2 v_1^2 &-g g_1 v_1^2 &0&0&...&0&0&0&0\\ -g g_1 v_1^2&
g_1^2 (v_1^2+v_2^2) & - g_1 g_2 v_2^2 &0 &...&0 &0&0&0\\ 0 & -g_1
g_2 v_2^2 & - g_2^2(v_2^2 + v_3^2) &- g_2 g_3 v_3^2 &...&0
&0&0&0\\ . & . & . &. &...&. &.&.&.\\  . & . & . &. &...&.
&.&.&.\\ 0 & 0 & 0 &0 &...& 0 & -g_{N-1} g_N v_{N-1}^2  & g_N^2
(v_{N-1}^2 + v_N^2)& g_N g' v_N^2\\0 & 0 & 0 &0 &...& 0 & 0 & g_N
g' v_N^2& g^{'2} v_N^2
\end{array} \right)\eeq}

Note that in the limit where $N=0$ this description of the Goldstone
modes of the model is simply the Standard Model. 
In fact to completely recover the Standard Model the higgs in 
the UV completion must also be made real. 

For larger $N$ when the couplings and vevs are all equal the W
mass matrix has eigenvalues \cite{csaki1}

\beq M^k_W = g v \sin \left[ {(2k-1)\pi \over 4N-2} \right] \eeq 
which for large $N$ and $k < N$ reproduces a KK like
tower of W states. Note that the W tower masses are suppressed
relative to $v$ by a factor of $N$. This is the mechanism
by which we will remove the higgs from the low energy spectrum.
In fact the couplings $g$ grow as $\sqrt{N}$ to keep the low
energy coupling invariant so the gain in higgs mass is only $\sqrt{N}$ too. 
In a simple higgs model the higgs mass is given by
$\sqrt{\lambda} v$ with $\lambda$ the quartic coupling in the higgs potential. 
Thus as the higgs vev increases by a factor of $\sqrt{N}$ so
does it's mass. In fact in the UV completion the scalar potential could 
be considerably more complicated with renormalizable terms of the form 
$|h_i|^2 |h_j|^2$ affecting the masses, but it is only our intention here 
to study the dependence of the vev on $N$ which is indicative of the higgs
mass scale. Unitarity in W scattering must still be maintained
at scales of order the lightest W mass - as discussed in \cite{sekhar1} the KK
tower acts as the restoring mechanism.

This simple set up will not make for good phenomenology since the
first KK partner of the W is very light (the direct experimental
bound is of order 500 GeV). We must therefore look at limits where
the KK modes are starting to become more massive and decouple.
There are two obvious limits of this form. Firstly we can raise
the vevs $v_1 - v_{N}$; in the limit where they are infinite the
low energy theory just becomes the Standard Model. This limit
seems promising since precision parameters will naturally tend to
the Standard Model values in this limit too (we will see soon how
well the new physics decouples). However, the lightest higgs
is becoming Standard Model like too in this limit and hence light.
The second limit, explored in \cite{schmidt}, is to take the couplings
$g_1-g_{N}$ to be large - this makes the KK modes heavy but does
not precisely return the Standard Model even in the infinite
coupling limit. Varying the vevs and couplings along the chain 
corresponds at the five dimensional level to warping the geometry 
\cite{sfetsos}
so we might hope to find the same successes seen in such models. 
We will explore both of these limits shortly.

To present results that can be compared to experimental data we
will numerically solve for the eigenvalues of the matrices 
(\ref{W},\ref{Z}). At this stage
we will simply work at tree level and search for a theory compatible
with the data at this level. We
must also couple the Standard Model matter fields into the
model. We will follow \cite{schmidt} and allow the fermions to couple to the
end two gauge groups in the moose chain. This choice ensures
that $T = 0$ \cite{schmidt} when the central
SU(2) groups' couplings are taken large. We have also explored 
other assignments but found little benefit from them. As usual we will
fix our model to the measured values of $M_Z$ the electric charge 
$e$ and the Fermi constant $G_F$
since these are the best measured experimental results.

Let us first study the limit where the gauge couplings of the
internal groups of the moose are taken large. We fix $g_1 ... g_n =
\tilde{g}$ and take couplings for the end groups $g$ and $g'$ as
shown in Fig 1. We also keep all the scalar vevs equal for now.
Fixing the parameters to the above experimental quantities is a little 
involved.  For a given value of $\tilde{g}$ the Z mass matrix depends 
on three parameters $g,g',v$ and determines the two parameters 
$M_Z, e$. Numerically we
first fix the physical value of $e$, which is given by 
$e = g \alpha_1 + g' \alpha_2$ where the
$\alpha$s are the relevant mixings, to fix $g$ as a function of $g'$.
The lowest mass eigenvalue of the Z mass matrix is then fixed to the 
physical Z mass  
giving us numerical values for  $g$ and $g'$ as a function of $v$.
Finally we diagonalize the W mass matrix and determine $v$ by requiring 
the physical value of $G_F$ which fixes each of $g,g',v$.

\begin{center}
\hskip-10pt{\lower15pt\hbox{ \epsfysize=3.5 truein
\epsfbox{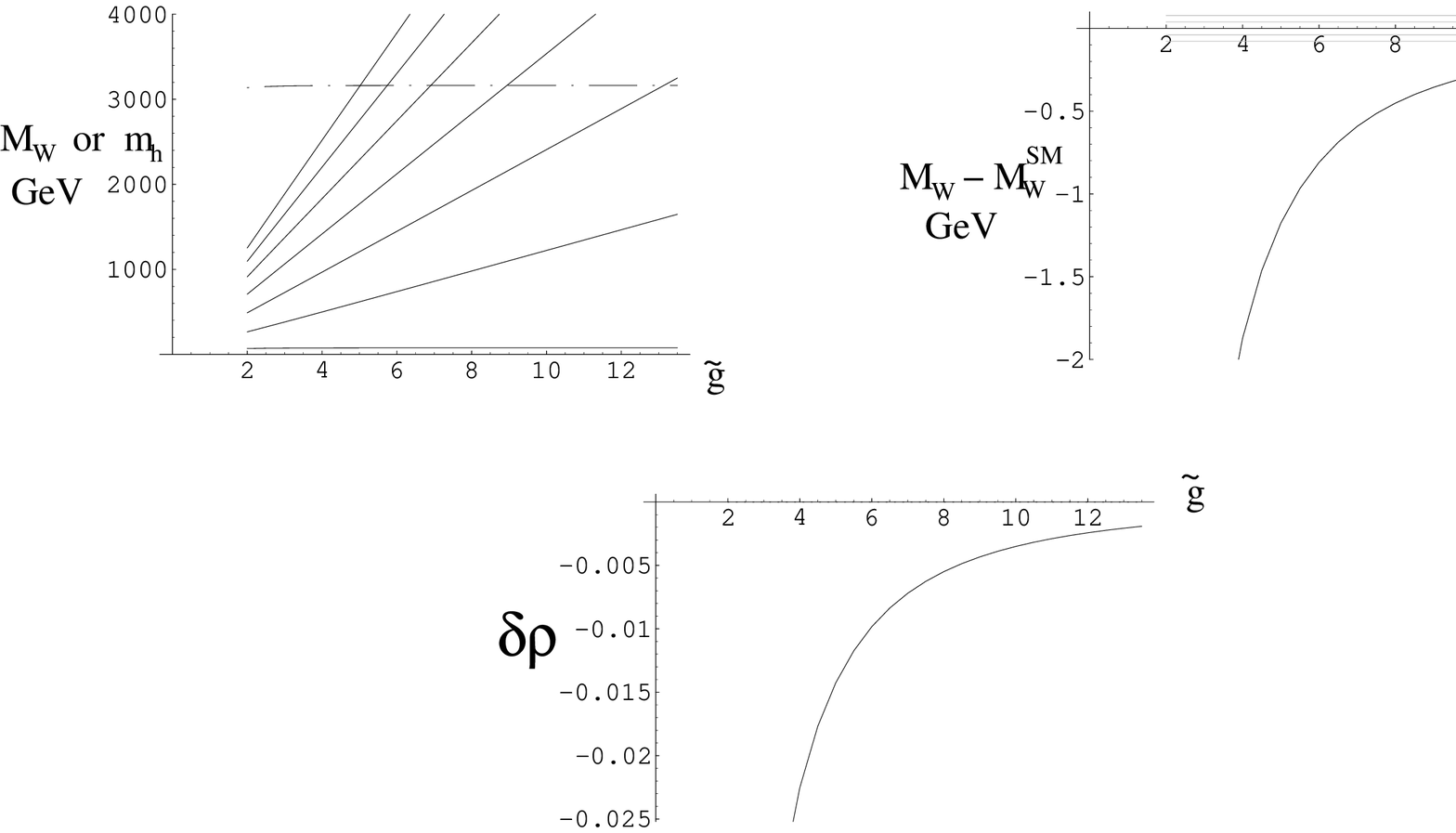}}}    \medskip\\[5mm] 
Figure 2: Results in the model with N=10. The first figure shows the 
tower of W boson states against changing $\tilde{g}$. The dot-dashed
line is the maximum higgs mass in the model. The second
figure shows how the lightest W boson mass deviates from the 
Standard Model value and the experimental constraints at 1 and 2 $\sigma$.
The third figure shows $\delta \rho$ in the model - the experimental 
constraints is $\delta \rho \geq -0.001$. 
\end{center}

We plot for the case where $N=10$ the W and KK W masses as a
function of $\tilde{g}$ in Fig 2. We also plot the maximum higgs
mass which we take to be a factor of $1TeV/(125)GeV$ times the
higgs vev (thus we would say the Standard Model higgs must lie
below 1TeV). The first success is that the higgs boson mass rises
by a factor of 3 relative to the Standard Model. The KK modes also
become more massive as  $\tilde{g}$ grows. However, the second figure showing
just the W mass and the experimental bounds reveals that the model
is some way off the data even as $\tilde{g}$ approaches $4 \pi$,
an absolute maxmum value for perturbativity. Finally we have also plotted
$\delta \rho = M_W^2/(M_Z^2 \cos \theta_w)-1$ with $\cos \theta_w$ obtained by 
equating the
electron coupling to the Z in our model to the Standard Model expression.
We find a negative value of $\delta \rho$ that falls as 
$\tilde{g} \rightarrow \infty$ but is still quite substantial at
$\tilde{g}= 4 \pi$. The relation between our definition of $\delta \rho$ and 
the parameter $T$ is discussed in more detail in the appendix.

It's interesting to look at the $N$ dependence of these models. As
an example we plot in Fig 3 the W mass as a function of $N$
when $\tilde{g} = 4 \pi$. We see that increasing $N$ actually
moves the results away from the data (although of course the rise
in the higgs mass grows as $\sqrt{N}$ so one would hope to strike
a balance).

\begin{center}
\hskip-10pt{\lower15pt\hbox{ \epsfysize=2. truein
\epsfbox{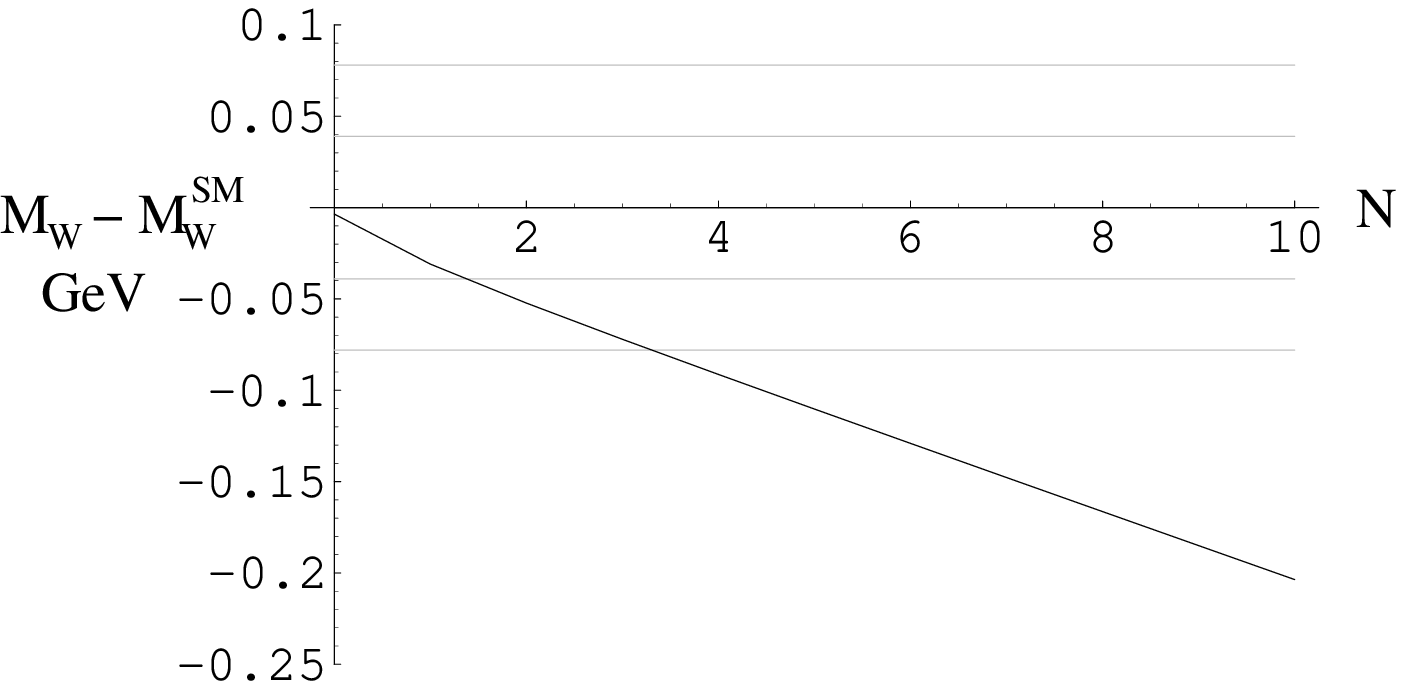}}} \medskip\\[5mm] 
Figure 3: The dependence of $M_W$ on $N$ in the moose models with
$\tilde{g}=4 \pi$.
\end{center}

We have though at our disposal a second decoupling limit for the
KK modes. Consider again the moose model with $N=10$.
If we take all the vevs except one to infinity then the
model returns to the Standard Model and all experimental
constraints will be met! As an example of this in Fig 4 we plot
the gauge boson masses in models with $\tilde{g}=2$ and $v_1$ to
$v_{N}= \tilde{v}$ and $v_{N+1}=v$. We plot the results for each value of
$\tilde{v}/v$. The extra decoupling is again apparent as the KK Ws rise
in mass. However, here we know that the higgs mass must fall to
the Standard Model value as $\tilde{v}/v \rightarrow \infty$ and from Fig 4 we
see that this fall in the higgs mass is much more dramatic than
the rise induced in the KK tower - precisely the opposite of what
one would require phenomenologically. 

\begin{center}
\hskip-10pt{\lower15pt\hbox{ \epsfysize=2. truein
\epsfbox{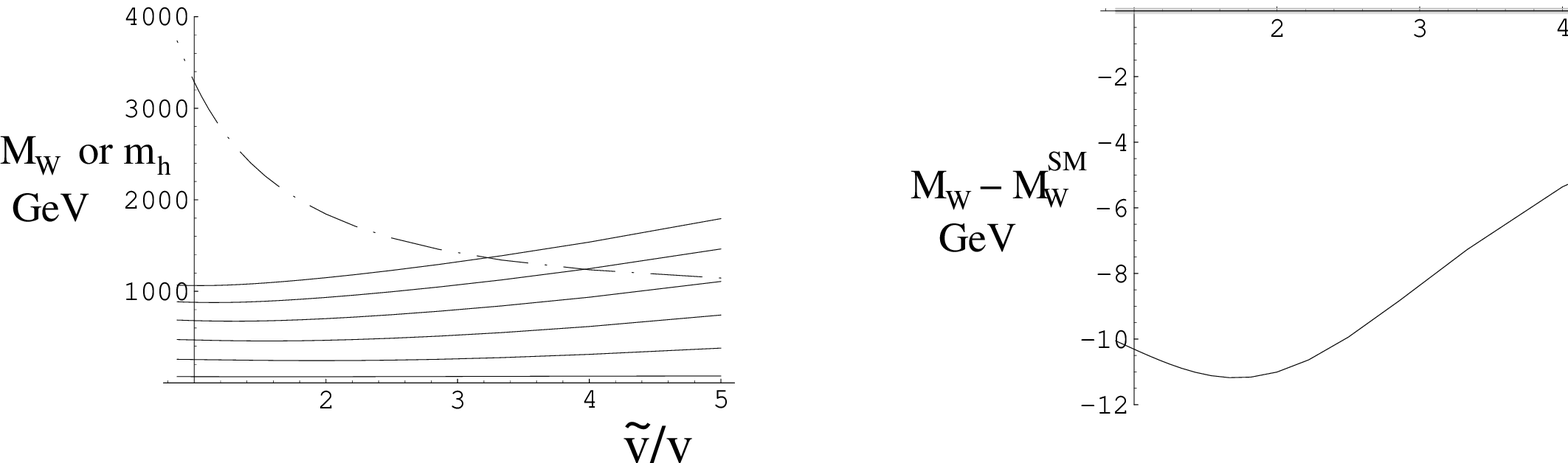}}}   \medskip\\[5mm] 
Figure 4: Results for the lightest Ws and higgs mass in the $N=10$ model with
$\tilde{g}=2$ and for varying the vevs $\tilde{v}/v$. The right-hand 
side of the plot approaches the Standard Model limit. 
\end{center}

Although these two limits are not themselves sufficient to provide
an interesting model one might hope that some combination of the
two scenarios might. As an example in the $N=10$ model we can set $\tilde{g}=
4 \pi$ and look at varying $\tilde{v}/v$. We plot the results in Fig 5
for $M_W$ and $m_h$.

\begin{center}
\hskip-10pt{\lower15pt\hbox{ \epsfysize=2. truein
\epsfbox{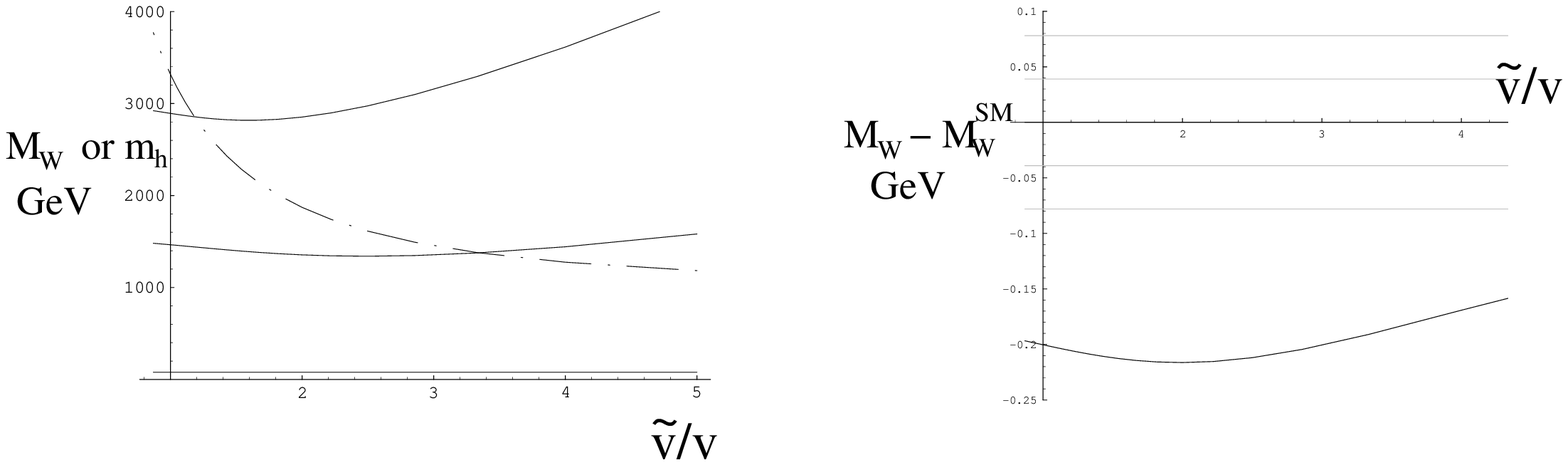}}}    \medskip\\[5mm] 
Figure 5: Results for the lightest Ws and higgs mass in the $N=10$ model with
$\tilde{g}=4 \pi$ and for varying the vevs $\tilde{v}/v$.
\end{center}

In fact in the area of parameter space where the higgs is heavy ($> 2 TeV$)
$M_W$ is still some way off the data. We have tried numerically adjusting the 
vevs and couplings along the chain but have not found any significant
improvement in the match to the data. In fact it is clear in this case that
most of the heavy Ws are so heavy they do not influence low energy unitarity.
For this reason it makes sense to restrict to models with small values of $N$
which we saw above are in better accord with data. We do though want to 
see a gain in the higgs mass which requires large N - as a compromise lets 
consider $N=4$ from which we can hope to obtain a rise in higgs mass of a 
factor of 2. We again set $\tilde{g}=4 \pi$ and plot results against $\tilde{v}
/ v$ in Fig 6.

Both $M_W$ and $\delta \rho$ lie closer to the data when the higgs is somewhat 
heavier than in the Standard Model in this case. In fact though the next
W in the tower of states lies close to 2 TeV in this model. This was not the 
intention for these models where it was hoped that the LHC would find heavy Ws 
and no higgs.
Here we have identified, at best, the possibility that a combination of a higgs
and quite strongly coupled W both at 2 TeV might viably restore unitarity
(a similar conclusion is reached in \cite{burdman} by an analysis of five 
dimensional models). 
The LHC might struggle to identify such a model. We conclude that finding 
a higgsless model in agreement with the precision data appears very hard - one
would perhaps need at least some subtle symmetries in a model to achieve 
this goal. \bigskip

\begin{center}
\hskip-10pt{\lower15pt\hbox{ \epsfysize=3.5 truein
\epsfbox{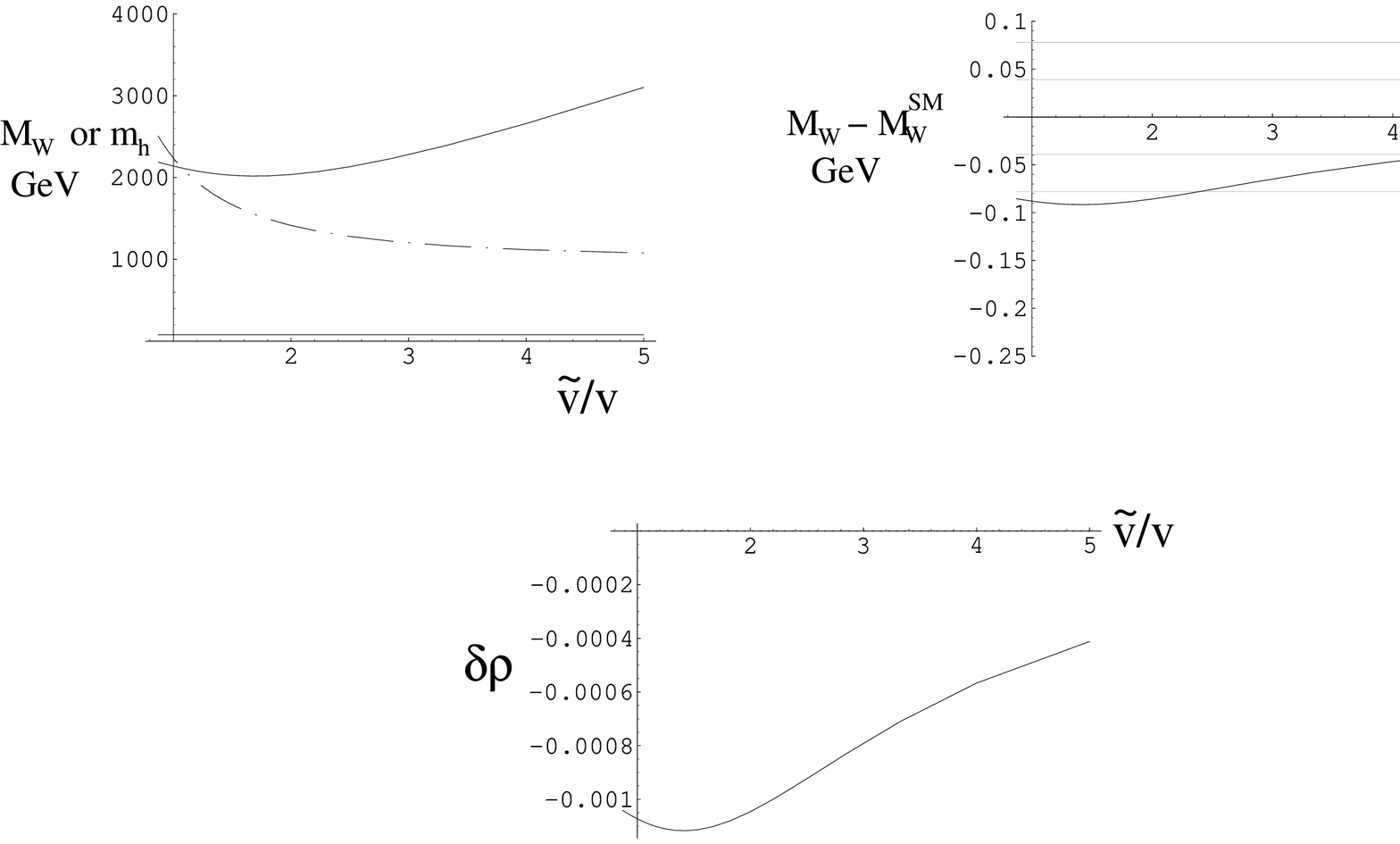}}} \medskip\\[5mm] Figure
6: Results for the lightest Ws and higgs mass in the $N=4$ model with
$\tilde{g}=4 \pi$ and for varying the vevs $\tilde{v}/v$. We also show
$\delta \rho$ vs $\tilde{v}/v$. Here the model
lies close to the experimental bounds with a higgs of order 2 TeV.
\end{center} \bigskip

\noindent {\bf Appendix}

Since the release of this work as a preprint the authors of \cite{sekapp}
have pointed out the following interesting relation between $\delta \rho$
as we define it and the parameters $S,T$. The models we study have been 
shown to have the parameters $T=U=0$ and positive S in \cite{kurachi}.
$\delta \rho$ can be defined either
as we do in terms of the pole W and Z masses or as the ratio of the
charged and neutral weak currents at low energies. This latter definition
gives $\delta \rho =  \alpha T$. The pole W and Z masses, given in terms of the
parameters $e^2, s^2, G_F$ and corrections 
$S,T,U$ are given by \cite{sekapp}

\beq M_Z^2 = {1 \over 4 \sqrt{2} G_f \left( {s^2 c^2 \over e^2} - 
{S \over 16 \pi} + s^2 c^2 {T \over 4 \pi} \right)}\eeq

\beq M_W^2 = {1 \over 4 \sqrt{2} G_f \left( {s^2  \over e^2} - 
{S + U \over 16 \pi} \right)}\eeq
So defining $\rho$ as $M_W^2/M_Z^2 c^2$ gives

\beq
\delta \rho = \alpha T - {\alpha S \over 4 c^2} + {\alpha U \over 4 s^2}
\eeq
Thus our negative values for $\delta \rho$ are consistent with and caused
by the presence of positive $S$.\bigskip

\noindent {\bf Acknowledgements:} We are grateful for discussions with 
Sekhar Chivukula, Liz Simmons and Tim Morris. 

\end{document}